\documentclass[aps,prb,twocolumn,floatfix]{revtex4-1}

\usepackage{amsmath,amssymb}
\usepackage{color}
\usepackage[pdftex]{hyperref,graphicx}
\hypersetup{colorlinks = true, urlcolor = blue, linkcolor = blue, citecolor = blue}
\usepackage{wscmath}

\providecommand{\cc}{c^{\phantom\dag}}
\providecommand{\cd}{c^\dag}

\begin{document}
\title{Novel magnetic state in $d^4$ Mott insulators}
\author{O. Nganba Meetei, William S. Cole, Mohit Randeria, Nandini Trivedi}
\affiliation{Department of Physics, The Ohio State University, Columbus, Ohio 43210, USA} 
\begin{abstract}
We show that the interplay of strong Hubbard interaction $U$ and spin-orbit coupling $\lambda$ in systems with $d^4$ electronic configuration leads to several unusual magnetic phases. Most notably, we find that competition between superexchange and spin-orbit coupling leads to a phase transition from a non-magnetic state predicted by atomic physics to a novel magnetic state in the large $U$ limit. We show that the local moment changes dramatically across this phase transition, 
challenging the conventional wisdom that local moments are 
robust against small perturbations in a Mott insulator. The Hund's coupling plays an important role in determining the nature of the magnetism. 
We identify candidate materials and present predictions for Resonant X-ray Scattering (RXS) signatures of the unusual magnetism in $d^4$ Mott insulators.
%
\end{abstract} 
\date{\today}
\maketitle

\section{Introduction}

Strong interactions lead to phenomena such as high $T_c$ superconductivity \cite{highTc_RMP_2006} and colossal magnetoresistance \cite{salamon_2001}. On the other hand, spin-orbit coupling (SOC) 
alone can lead to topological band insulators \cite{hasan_2010}. These two features naturally combine in the $4d/5d$ transition metal materials, which hold the potential of hosting new phases of matter with entangled spin, orbital and charge degrees of freedom. Already there are many predictions for exotic topological matter, for example the topological Mott insulators \cite{pesin_2010} and Weyl semi-metals \cite{vishwanath_2011}. Recent experiments demonstrating that Sr$_2$IrO$_4$ is an unusual Mott insulator with a half filled  $J=1/2$ band resulting from strong SOC \cite{kim_2009} have prompted the search for Weyl semi-metals in iridium pyrochlores \cite{ueda_2012}.

\begin{figure}[t]
\includegraphics[width=0.48\textwidth]{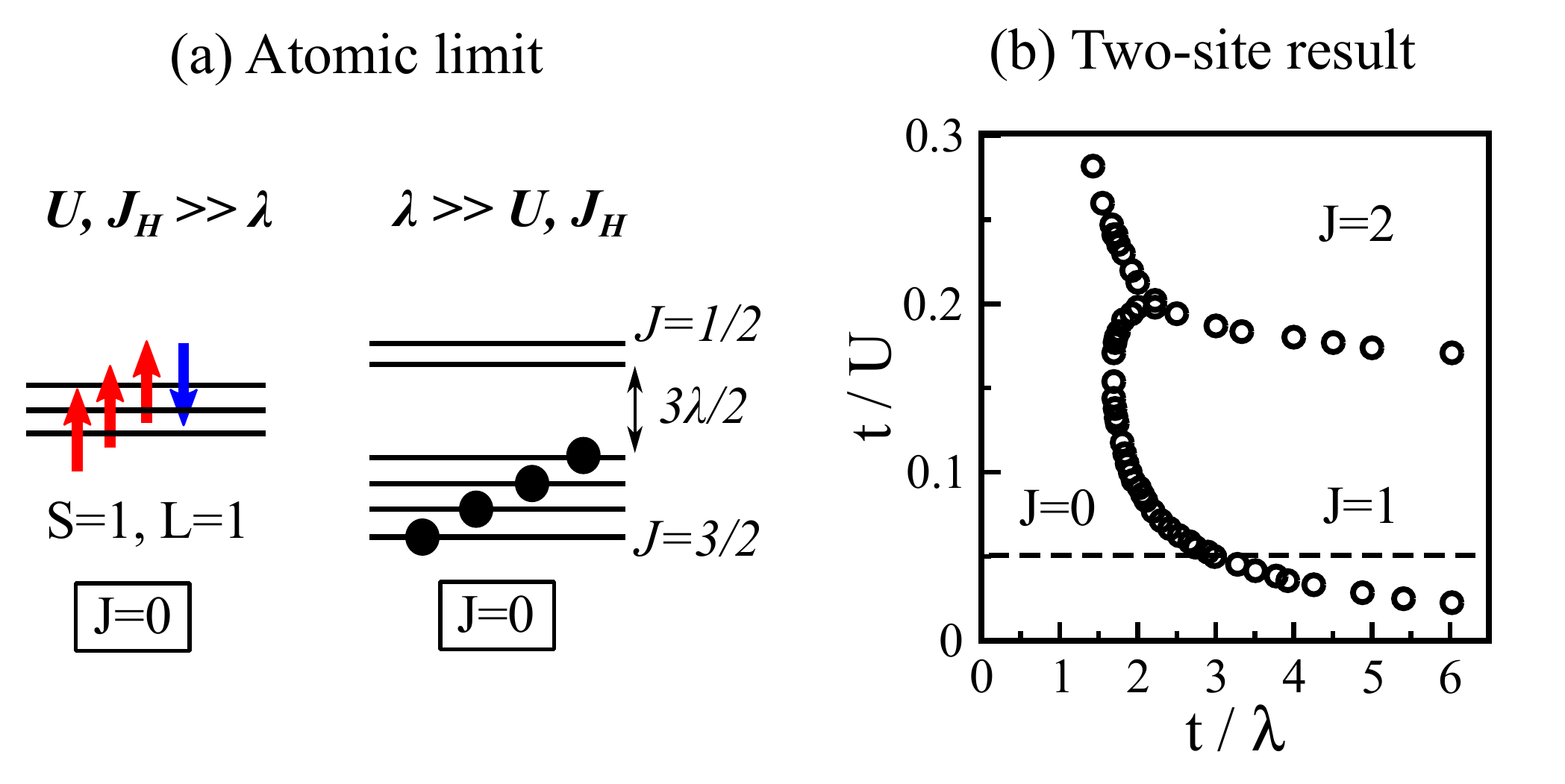}
\caption{(a) The atomic ground state of $d^4$ ions in both $U, J_H \gg \lambda$ and $\lambda \gg U,J_H$ limits is non-magnetic with $J=0$.  (b) The two-site phase diagram of the $d^4$ system calculated by exact diagonalization shows the existence of ferromagnetic phases ($J=1$ and $J=2$) in addition to the non-magnetic ($J=0$) phase in the $U-\lambda$ plane. We have used $J_H=0.2U$, the typical value for 4d/5d oxides. Dashed line indicates $U=20t$ relevant for candidate materials proposed here \cite{footnote}.}
\label{Fig:1}
\end{figure}

Most of the focus in this field to date has been on iridium based materials with a $d^5$ electronic configuration that can be understood in terms of a half-filled $J=1/2$ manifold arising from large spin-orbit induced splitting of $t_{2g}$ orbitals. The physics is dramatically different for other fillings. Mott insulators with $d^1$ and $d^2$ configurations have been shown to exhibit exotic magnetic phases \cite{chen_2010,chen_2011} in the presence of large SOC.  In the $d^3$ case, SOC is quenched in a cubic environment \cite{meetei_2013} and the problem reduces to a conventional spin-only model. This leaves  the $d^4$ case, which has been largely ignored because large SOC and strong interactions is expected to give rise to a non-magnetic state in the atomic limit \cite{chen_2011} (see Fig.~\ref{Fig:1}(a)).


We show that, contrary to naive expectations, the $d^4$ configuration has a rich magnetic phase diagram as a function of SOC, Hubbard $U$, and
Hund's coupling $J_H$. 
For large $U$ in particular, the atomic limit is a non-magnetic insulator of local $J=0$ singlets. Turning on hopping   
leads to two unusual phenomena: (a) quantum phase transition from the expected non-magnetic insulator to a novel magnetic state, (b) local moments are spontaneously generated in the magnetic phase due to superexchange-induced mixing of the on-site singlet state with higher energy triplet states.
We emphasize the importance of Hund's coupling $J_H$ in determining the sign of the superexchange interaction between these local moments. 
If $J_H$ is ignored, the superexchange is antiferromagnetic \cite{khaliullin_2013}, however we show that even a modest value of the Hund's coupling,
which is realistic for 4d and 5d transition metal oxides, leads to a ferromagnetic interaction between moments. 

Our result provides a counterexample to the commonly held notion that Mott insulators have well defined local moments that cannot be affected by perturbations that are small compared with the interaction scale $U$. We further present predictions for resonant x-ray scattering (RXS). Unlike the iridates with $d^5$ configuration, the RXS amplitude in $d^4$ Mott insulators depends on the strength of SOC. We conclude by identifying candidate materials among the ruthenates. %

\section{Hamiltonian}

We consider a three-orbital Hubbard model with SOC which captures the essence of $4d$/$5d$ materials. Cubic crystal field splitting is typically larger than  $U$ \cite{lee_2003,biermann_2012}, and for $d^4$ configuration only the $t_{2g}$ orbitals are occupied. The situation is reversed in the case of $3d$ materials \cite{lee_2003,imada_1998} and the $e_g$ orbitals also come into play. The Hamiltonian under consideration is given by: 
\begin{equation}
H = H_{hop} + \sum_i \left( H_{i,U} + H_{i,SOC} \right) \label{eq:hamiltonian}
\end{equation}

where 
\begin{eqnarray}
 H_{hop} &=& \sum_{ij} \sum_{\alpha\beta} \sum_{\sigma\sigma'} \left( t_{ij}^{\alpha\sigma,\beta\sigma'} \cd_{i\alpha\sigma}\cc_{j\beta\sigma'} + \mbox{h.c.} \right) \\
   H_{i,SOC} &=& \lambda \sum_{\alpha\beta} \sum_{\sigma\sigma'} \langle\mathbf{s}\cdot\mathbf{l}\rangle_{\alpha\sigma,\beta\sigma'}\cd_{i\alpha\sigma}\cc_{i\beta\sigma'}  \\
   H_{i,U} &=& (U-3J_H)\frac{\hat{N}_i(\hat{N}_i-1)}{2} + \frac{5}{2}\hat{N}_i  \nonumber \\
           &-&  2J_H \hat{S}^2_i  - \frac{1}{2}J_H \hat{L}^2_i 
\end{eqnarray}
Here $\cd_{i\alpha\sigma}(\cc_{i\alpha\sigma})$ creates(annihilates) an electron at site $i$ in orbital $\alpha$ with spin $\sigma$. $\hat{N}_i$, $\hat{S}_i$ and $\hat{L}_i$ are the total occupation number, total spin momentum and total orbital momentum operators at site $i$.  $t_{ij}^{\alpha\sigma,\beta\sigma'}$ is the hopping matrix element from the state $\beta\sigma'$ at site $j$ to $\alpha\sigma$ at site $i$. We consider only nearest neighbor hopping, and take it to be diagonal in both spin and orbital space ($t_{ij}^{\alpha\sigma,\beta\sigma'}\rightarrow t_{ij}\delta_{\alpha\beta}\delta_{\sigma\sigma'} $). This symmetry allows a more transparent understanding of the exact diagonalization results, but does not effect the qualitative features of the low energy physics, compared to a more realistic choice of $t_{ij}^{\alpha\sigma,\beta\sigma'}$ (see Appendix \ref{appen:D} for more details). $U$, $J_H$ and $\lambda$ are intra-orbital interaction strength, Hund's coupling and SOC strength respectively. $\langle\mathbf{s}\cdot\mathbf{l}\rangle_{\alpha\sigma,\beta\sigma'}$ are the matrix elements of atomic SOC in the $t_{2g}$ basis. Note that the $t_{2g}$ orbitals have an effective orbital momentum $l=1$ with opposite sign of SOC. All energy scales are measured in units of $t$. 


\begin{figure}[t]
\includegraphics[width=0.45\textwidth]{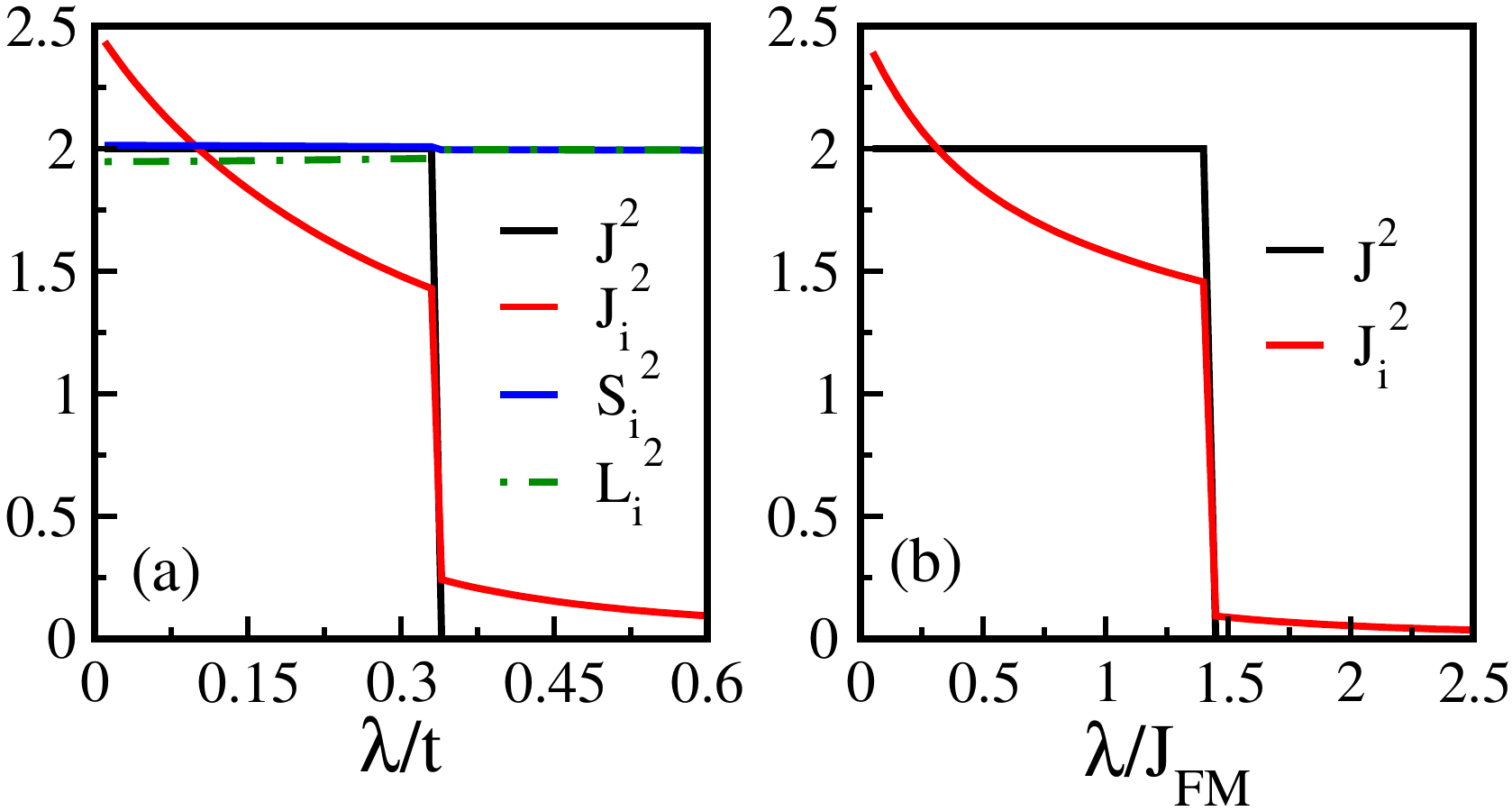}
\caption{(a) Two-site exact diagonalization result of the full electronic Hamiltonian in Eq.~\ref{eq:hamiltonian} showing magnetic phase transition as a function of $\lambda$ in the Mott limit ($U/t=20$ and $J_H=0.2U$). The total $J$-moment changes from $J=1$ to $J=0$ at $\lambda_c/t\approx 0.33$. The \emph{local} moment $J_i$ shows a discontinuous jump at the phase transition, while the local $S_i$ and $L_i$ moments do not change. (b) The low energy magnetic Hamiltonian in Eq.~\ref{eq:spin_hamiltonian} accurately captures the phase transition from $J=0$ to $J=1$ with decreasing $\lambda$ at the two site level.}
\label{Fig:2}
\end{figure}

\section{Two-site results} 

We first present exact results obtained by numerical diagonalization of Eq.~\ref{eq:hamiltonian} for a two-site system with 8 electrons defining a Hilbert space of $^{12}C_8=495$ basis states. 
We have used $J_H=0.2 U$ relevant for 4d/5d oxides \cite{van_der_marel_sawatzky_1988,georges_2013}.
The two-site system exhibits three different magnetic states as a function of $\lambda$ and $U$ as shown in Fig.~\ref{Fig:1}(b): (i) a non-magnetic state ($J=0$) in the large $\lambda$ limit, (ii) a ferromagnet with $J=2$ for small $\lambda$ and moderate $U$ and (iii) a ferromagnet with $J=1$ at large values of $U$ and small $\lambda$. Here $J$ refers to the total moment in the ground state of the two-site system. While $J\neq 0$ is a good diagnostic for magnetic states even on a lattice, the integer values in Fig.~\ref{Fig:1}(b) are specific to the two site system.  Strictly at $\lambda=0$ all $J$ states become degenerate and the states should be labeled in terms of spin ($S$) and orbital ($L$) moments.

The two site results are most reliable in the Mott limit with large $U/t$ where charges are spatially localized and the single particle excitation gap is set by $U$. In this limit, the atomic picture is expected to give a good description of local properties. For $d^4$ Mott insulators, the atomic picture would predict decoupled non-magnetic ions at every site (see Fig.~\ref{Fig:1}(a)). Instead, we find a phase transition from a total $J=0$ state to a total $J=1$ state with decreasing $\lambda/t$ (see Fig.~\ref{Fig:1}(b)). The origin of this behavior lies in a dramatic change in the expectation value of the local moment $\langle J^2_i \rangle$ across the magnetic phase transition as shown in Fig.~\ref{Fig:2}(a). In a conventional Mott insulator, the local moment is determined by the large interaction scale $U$ and any perturbation which is small compared to $U$ does not affect the local moment. In the case of $d^4$ Mott insulators, $U$ fixes the local $S_i=1$ and $L_i=1$ moments which are robust as shown in Fig.~\ref{Fig:2}(a). The total moment $J_i$, on the other hand, is determined by $\lambda$ which is a small parameter. As we will show later, super exchange interaction competes with $\lambda$ and gives rise to the magnetic phase transition.  

Away from the large $U$ limit, the two site calculation is less reliable. However, we can easily understand the limiting cases.  The $J=0$ state at large $\lambda$ and small $U$ corresponds to a band insulator with a completely filled $J=3/2$ manifold. It is smoothly connected to the $J=0$ Mott insulator at larger $U$, consistent with recent Gutzwiller and Dynamical Mean-Field Theory calculations \cite{du2_2013}. The post-perovskite material NaIrO$_3$ and perovskites BaOsO$_3$ and CaOsO$_3$ are believed to be in such a state \cite{du1_2013,du2_2013}. In the limit of small $\lambda$ and moderate $U$, the $J=2$ ferromagnet is essentially the Stoner ferromagnet seen in SrRuO$_3$ \cite{mazin_1997}.

\begin{figure*}[t]
\includegraphics[width=0.87\textwidth]{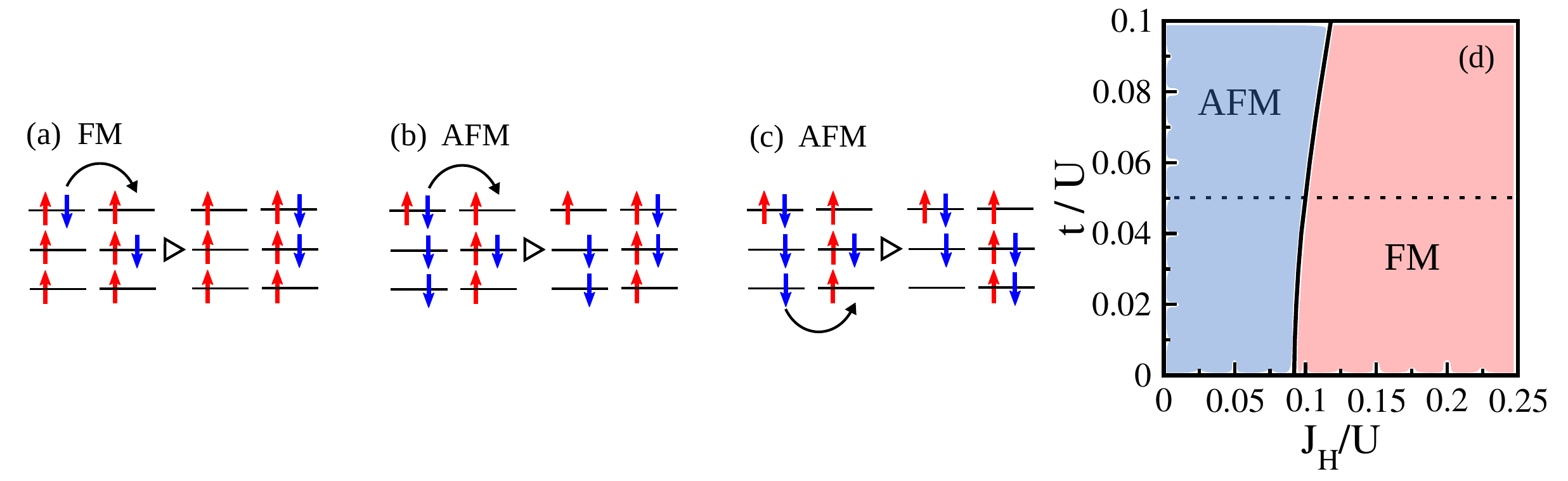}
\caption{Superexchange pathways: (a) ferromagnetic, (b) and (c) antiferromagnetic. Hund's coupling favors the ferromagnetic pathway. (d) Exact diagonalization result of the electronic Hamiltonian in Eq.~\ref{eq:hamiltonian} with anisotropic hopping relevant for $t_{2g}$ orbitals showing the sign of superexchange interaction as a function of $t/U$ and $J_H/U$. Dashed line indicates $U=20t$ \cite{footnote}. }
\label{Fig:3}
\end{figure*}

\section{Role of Hund's coupling}

The sign of the superexchange interaction in $d^4$ Mott insulators is greatly affected by Hund's coupling $J_H$. Starting with an AFM superexchange 
for $J_H=0$, in agreement with Ref.~\cite{khaliullin_2013}, we show from the exact diagonalization of a two-site problem, there is a phase transition to a FM superexchange (see Fig.~\ref{Fig:3}(d)).
Specifically, for realistic   parameters of $U\approx 20t$ \nocite{biermann_2012}\cite{footnote} and  $J_H\approx 0.2 U$ \cite{van_der_marel_sawatzky_1988, georges_2013} relevant for $4d$ oxides, the superexchange is firmly in the ferromagnetic regime. 

To gain insight into the exact result in Fig.~\ref{Fig:3}(d), we have also done a simplified perturbative calculation using the ferromagnetic (FM) and anti-ferromagnetic (AFM) states as shown in Fig.~\ref{Fig:3}(a)-(c).  With $J_H=xU$ the energy gained by the FM state is $ \Delta E^{FM} = -\frac{2t^2}{U}\frac{1}{1-3x} $ and by the AFM state is $ \Delta E^{AFM} = -\frac{2t^2}{U}\left(\frac{1}{3(1-3x)} + \frac{7}{6} + \frac{1}{2(1+2x)}\right) $ (details in Appendix \ref{appen:A}). For $x=0.2$ our perturbative analysis also gives rise to a ferromagnetic superexchange.

\section{RXS cross-section}

We now make predictions for RXS cross-sections, which can be used to characterize the ferromagnetic insulator. For Mott insulators, RXS matrix elements are usually calculated in the free ion approximation\cite{fink_2013,kim_2009}. However, to include non-local effects, which we show is crucial for understanding the $d^4$ ferromagnetic Mott insulator, we need to generalize the expression for RXS amplitude as follows (see Appendix \ref{appen:C})
\begin{equation}\label{eq:RXS}
\Delta f(\omega) \propto \mbox{Tr} \left[ \rho \sum_{n}\frac{ (\mathbf{e'}.\mathbf{D})^\dagger \lvert \psi_n\rangle\langle \psi_n \rvert \mathbf{e}.\mathbf{D} }{ E_n - E_0 - \hbar \omega -i\Gamma_n } \right]
\end{equation}
where $\rho$ is the \emph{reduced} density matrix at the scattering site, and the trace is over atomic states in the $d^4$ configuration. Here $\mathbf{e}$($\mathbf{e'}$) is the incoming(outgoing) polarization, $\mathbf{D}$ is the dipole operator, $\lvert \psi_n \rangle$ is an excited state (in the $d^5$ configuration) with energy $E_n$, and $E_0$ is the ground state energy. $\Gamma_n$ is the inverse life time of the excited state $\lvert \psi_n \rangle$.

\begin{figure}[ht]
\centering
\includegraphics[width=0.48\textwidth]{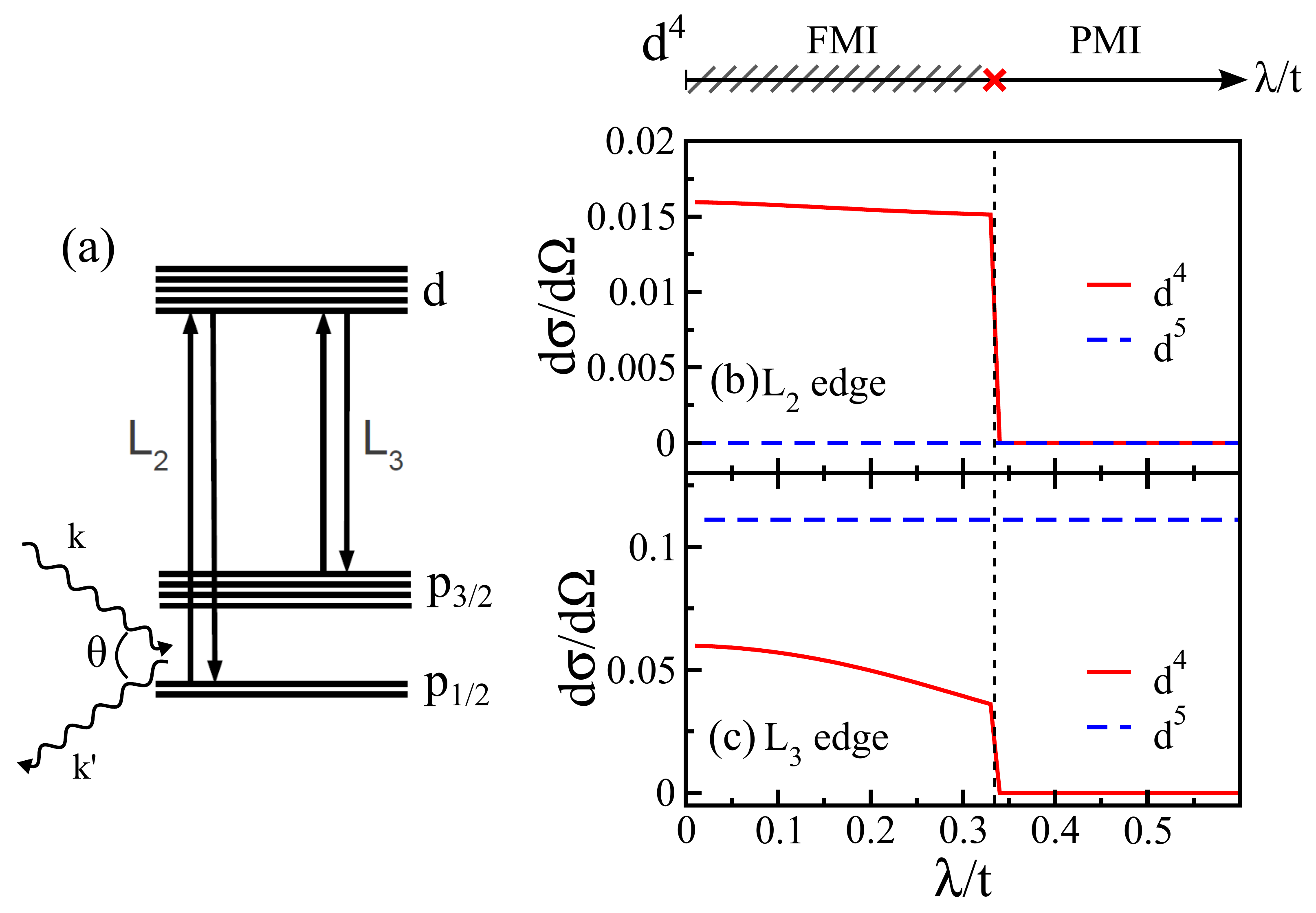}
\caption{(a) Schematic of resonant X-ray elastic scattering (RXS) from momentum $\mathbf{k}$ to $\mathbf{k}'$ with scattering angle $\theta$. (b) and (c) Magnetic RXS cross-section for $L_2$ and $L_3$ edges respectively at $\theta=\pi/2$. For $d^4$ Mott insulators, the
scattering cross-section is identically zero in the paramagnetic insulator (PMI) at large $\lambda/t$ and undergoes a discontinuous jump at a critical $\lambda/t$ to a non-zero value in the ferromagnetic insulator (FMI). In sharp contrast, $d^5$ Mott insulators with half-filled $J=1/2$ manifold show no change with $\lambda/t$.}
\label{Fig:4}
\end{figure}

The $L_2$ and $L_3$ edges corresponds to excitations from $2p_{1/2}$ and $2p_{3/2}$ levels respectively to the intermediate $d^5$ states as shown in Fig.~\ref{Fig:4}(a). We ignore the SOC induced energy splitting among the $d^5$ states as it is much smaller than the inverse life time $\Gamma_n$, and consider resonant enhancement coming from all intermediate states.

The magnetic ($\sigma$-$\pi$) scattering cross-section at scattering angle $\theta=\pi/2$ for $d^4$ Mott insulators at the $L_2$ and $L_3$ edges are shown in Fig.~\ref{Fig:4}(b) and (c) respectively as a function of $\lambda/t$. The RXS matrix elements are calculated using the eigenstates of the two-site system in order to include the effects of super exchange. For comparison we have also included the results for $d^5$ Mott insulators, in which case atomic calculations suffice. As seen clearly from Fig.~\ref{Fig:4}(b) and (c), the resonant enhancement in scattering for $d^4$ Mott insulators changes with $\lambda/t$. This indicates the dependence of local physics on the competition between $\lambda$ and $J_{FM}\sim \mathcal{O}(t^2/U)$. In the non-magnetic state, $\lambda$ dominates local physics. Only the lowest energy $J_i=0$ state contributes to $\rho$ and magnetic RXS cross-section is identically zero. With decreasing $\lambda/t$, the systems becomes ferromagnetic and the higher energy local $J_i=1$ state also contributes significantly to $\rho$, resulting in non-zero magnetic scattering which depends on $\lambda/t$. In sharp contrast, the cross-sections for the $d^5$ case are independent of $\lambda/t$ and only the $L_3$ edge is resonantly enhanced. Note that an antiferromagnet in the absence of canting will not give rise to magnetic RXS scattering because the large X-ray spot size averages over local moments to give a zero net moment. Our RXS result can easily distinguish between the novel ferromagnetic insulator presented here and the anti-ferromagnetic insulator proposed in Ref.~\cite{khaliullin_2013}.


\section{Magnetic Hamiltonian and Ginzburg-Landau theory}   

Building on the intuition from the two-site results, we now extend our analysis to a lattice.  Using perturbation theory, we derive the following magnetic Hamiltonian  (see Appendix \ref{appen:B})
\begin{equation} 
\tilde{H} =-\frac{J_{FM}}{2}\sum_{\langle ij\rangle}\mathbf{S}_i\cdot\mathbf{S}_j \mathcal{P}(\mathbf{L}_i+\mathbf{L}_j=1)
+ \frac{\lambda}{2}\sum_i\mathbf{L}_i\cdot\mathbf{S}_i \label{eq:spin_hamiltonian}
\end{equation}
which includes SOC and superexchange mediated by only the lowest-lying virtual state. Here $S_i=1$ and $L_i=1$ are local $S$ and $L$ moments at site $i$, and $J_{FM}=4t^2/(U-3J_H)$  sets the ferromagnetic superexchange scale.  In order to maximize energy gain from virtual hops, each bond is projected by $\mathcal{P}(\mathbf{L}_i+\mathbf{L}_j=1)$ on to the total $L=1$ space.  $\mathcal{P}(\mathbf{L}_i+\mathbf{L}_j=1)$ also generates orbital entanglement which is unusual in the ferromagnetic state.  The factor of $1/2$ in the SOC term comes from rewriting the SOC Hamiltonian in the $L$-$S$ coupling scheme relevant for the $d^4$ configuration. The competition between $J_{FM}$ and $\lambda$ is clear in Eq.~\ref{eq:spin_hamiltonian}: the first term likes each bond to have $S=2$ and $L=1$, while the second term prefers each site to have $J_i=0$. As shown in Fig.~\ref{Fig:2}(b), the effective Hamiltonian in Eq.~\ref{eq:spin_hamiltonian} accurately reproduces the phase transition at the two-site level from a total $J=0$ to a $J=1$ state as a function of $\lambda/J_{FM}$.



On a lattice, the magnetic phase transition is best described in terms of bosonic operators $s_i^\dagger$ which creates a singlet at site $i$ and $T_{i,(0,\pm)}^\dagger$ which creates a triplet carrying $J_i^z=0,\pm$1 at site $i$ \cite{sachdev_1990,giamarchi_2008}. We ignore the $J_i=2$ states as they are much higher in energy.  By calculating the matrix elements of $S$ and $L$ operators in the singlet-triplet space, we get   
\begin{eqnarray} \label{eq:bond_operator}
S_i^\alpha &=& -\sqrt{\frac{2}{3}}\left( T_{i\alpha}^\dagger s_i + s^\dagger_i T_{i\alpha} \right) - \frac{i}{2}\epsilon_{\alpha\beta\gamma}T^\dagger_{i\beta}T_{i\gamma} \nonumber \\
L_i^\alpha &=& \sqrt{\frac{2}{3}}\left( T_{i\alpha}^\dagger s_i + s^\dagger_i T_{i\alpha} \right) - \frac{i}{2}\epsilon_{\alpha\beta\gamma}T^\dagger_{i\beta}T_{i\gamma} 
\end{eqnarray}
where $\alpha,\beta,\gamma=x,y$ or $z$, $T^\dagger_{iz}=T^\dagger_{i0}$, $T^\dagger_{ix}=-(T^\dagger_{i1}-T^\dagger_{i-1})/\sqrt{2}$ and $T^\dagger_{iy}=i(T^\dagger_{i1}+T^\dagger_{i-1})/\sqrt{2}$. Substituting these expressions into Eq.~\ref{eq:spin_hamiltonian}, we obtain the Hamiltonian in terms of bosonic operators.  Within the saddle-point approximation, the non-magnetic ground state which consists of singlets at every site is described by a condensate of singlets with a gap to triplet excitation. With increasing $J_{FM}$, the gap is reduced. The phase transition to the ferromagnetic state, described by a condensate of triplets, is signaled by the closing of the gap. Close to the phase transition, we assume $\langle s_i \rangle \approx 1$ and $\langle T_{i\alpha} \rangle=\phi_{i\alpha}  \ll 1$ and the effective Ginzburg-Landau functional with terms upto second order in $\phi_{i\alpha}$ is given by
\begin{eqnarray}
\mathcal{L} &=& \frac{\lambda}{2}\sum_{i\alpha}\left[\phi^*_{i\alpha} \quad \phi_{i\alpha}\right]
\left[ 
\begin{array}{cc}
1 & 0 \\
0 & 1
\end{array}
\right] 
\left[
\begin{array}{c}
\phi_{i\alpha} \\
\phi^*_{i\alpha}
\end{array}
\right] \nonumber \\
&& -  \eta J_{FM} \sum_{\langle ij \rangle,\alpha} \left[ \phi^*_{i\alpha} \quad \phi_{i\alpha}\right]
\left[
\begin{array}{cc}
1 & a \\
a & 1
\end{array}
\right] 
\left[
\begin{array}{c}
\phi_{j\alpha} \\
\phi^*_{j\alpha}
\end{array}
\right]
\end{eqnarray}
where $\eta$ and $a$ are parameters of $\mathcal{O}(1)$ which depend on details of the model. This can be solved easily by a Bogoliubov transformation which gives a gap function $\Delta_k=\sqrt{(\lambda/2 - \eta J_{FM}f(\mathbf{k}))^2 - (a\eta J_{FM} f(\mathbf{k}))^2}$ where $f(\mathbf{k})=\sum_\delta \text{cos}(\mathbf{k}.\mathbf{\delta})$ and $\mathbf{\delta}$ is nearest-neighbor position. The gap closes at $\mathbf{k}=0$ (indicating ferromagnetic phase) when $\lambda_c/J_{FM}=2z\eta (1+\lvert a \rvert)$ or $\lambda_c \sim z\mathcal{O}(t^2/U)$ where $z$ is the coordination number.

\section{Materials} 

We propose candidate materials from the double perovskite family ($A_2BB'O_6$, $A$ is an alkaline earth, $B$, $B'$  are two different transition metal ions, ordered in a 3D chequerboard pattern) which can be tuned across the magnetic transition by chemical substitution and/or pressure. If we choose the $B$  sites to have completely filled shells, and $B^\prime$ to be an active magnetic 4d/5d element, then the bandwidth is suppressed and SOC competes with $J_FM$ in the Mott insulating state. Of particular interest to us is La$_2$ZnRuO$_6$ which is an insulator with Ru in $d^4$  configuration. Two different samples grown by two different groups have shown different magnetic states; one group found a ferromagnetic state with $T_C\approx 165K$ \cite{dass_2004}, while the other found a non-magnetic state \cite{yoshii_2006} indicating that La$_2$ZnRuO$_6$ is close to the phase boundary so that small differences in the lattice parameter could produce this discrepancy. Another closely related material La$_2$MgRuO$_6$ \cite{dass_2004} is also a promising candidate. An RXS study under pressure will be an ideal experiment to observe the phase transition. 



\section{Conclusion}

In conclusion, coming from $3d$ oxides, the standard paradigm for Mott insulators is the following: (a) Local moments, determined by the large interaction scale $U$, are robust. (b) Atomic physics gives a good description of local properties. A naive extension to the $4d/5d$ oxides ions with four electrons in the $t_{2g}$ orbitals and with spin orbit coupling predicts non-magnetic J=0 singlets. Here we find a major departure from the standard paradigm for the $d^4$ case: (1) Local moments are no longer robust. Weak tunneling of electrons between atoms generates a local moment, and therefore, atomic physics is no longer adequate to describe the local properties. (2) The local moments once formed interact by the superexchange mechanism ($J_{FM}$) which depends crucially on Hund's coupling $J_H$. 
Hund's coupling favors ferromagnetic superexchange and for the $4d/5d$ oxides with typical $J_H/U\approx 0.2$ we predict a novel orbitally entangled ferromagnetic Mott insulator with distinct signatures in RXS scattering. Recent dynamical mean-field theory \cite{antoine_2011} and exact \cite{li_2014} results emphasized the role of Hund's coupling in electronic and magnetic properties, and our work adds another prime example of how Hund's coupling, which is often relegated to a secondary role compared to $U$, can be the driving force for novel magnetic states. While we have focused mostly on the physics at large $U$ and $\lambda$, the phase diagram in Fig.~\ref{Fig:1}(b) is extremely rich, allowing for a broader exploration of magnetic and metal-insulator phase transitions.

\section*{Acknowledgements}
We thank Daniel Khomskii and Patrick Woodward for fruitful discussions. We acknowledge support from the CEM, an NSF MRSEC, under Grants DMR-0820414 and DMR-1420451 (O.N.M., M.R., and N.T.) and from DOE BES DE-FG02- 07ER46423 (W.S.C.).

\appendix

\section{Role of Hund's coupling in Superexchange interaction} \label{appen:A}
Here we present the details of the simplied perturbative analysis discussed in the main text to determine the nature of superexchange interaction in $d^4$ Mott insulators. We use simple classical FM and AFM states shown in Fig.~\ref{Fig:3}(a,b,c) of main text, to get a clear intuitive picture of the role of Hund's coupling in favoring FM superexchange over AFM. A more accurate perturbative calculation would use quantum $S = 2$ (FM) and $S = 0$ (AFM) states that have entanglement in spin and/or orbital space. For instance, one needs a superposition of states with total L=1. We do not pursue this elaborate approach here since our only goal is to gain insight into the exact results of Fig.~\ref{Fig:3}(d). We also ignore spin-orbit coupling (SOC) in this analysis, which is later added back in the effective magnetic Hamiltonian within an L-S coupling scheme.  

In the atomic limit, the Hamiltonian for the $t_{2g}$ orbitals is 
\begin{eqnarray}\label{eq:H_at}
H_0 &=& \sum_i H^{at}_i  \\
H^{at}_i &=& \frac{(U-3J_H)}{2}\hat{N}_i(\hat{N}_i-1) +\frac{5}{2}J_H \hat{N}_i \nonumber \\
         && - 2J_H \hat{S}^2_i - \frac{1}{2}J_H \hat{L}^2_i
\end{eqnarray}
where $\hat{N}_i$, $\hat{S}_i$ and $\hat{L}_i$ are the number of electrons, total spin moment and total orbital moment operators at site $i$ respectively. They are all good quantum number in the atomic limit. We have ignored the chemical potential because we will fix the filling at $d^4$, for which the chemical potential term is just a constant.  

To analyze the superexchange interaction, we consider a two site system. The ground state in the atomic limit is in the $d^4-d^4$ configuration with $L_i=1$ and $S_i=1$.  Its energy is calculated easily using Eq.~\ref{eq:H_at}
\begin{equation}\label{eq:E0}
E_0 = 12U - 26J_H
\end{equation}

The hopping term acts as a perturbation to $H_0$ and is given by 
\begin{equation}\label{eq:kinetic}
H_{hop}=-t\sum_{\alpha,\sigma}(c^\dagger_{1\alpha\sigma}c_{2\alpha\sigma}+h.c.)
\end{equation}

The ground state gains energy via virtual hops and to determine the nature of the superexchange interaction generated by such virtual hops we calculate the energy gained by both ferromagnetic and anti-ferromagnetic states. Fig.~\ref{Fig:3}(a) of main text shows the ferromagnetic path way. The intermediate state in $d^3-d^5$ configuration has $S_1=3/2$, $L_1=0$, $S_2=1/2$ and $L_2=1$. Then using Eq.~\ref{eq:H_at}, the energy of the intermediate state is $E^{FM}_1 = 13U - 29J_H$. Then the energy gained by the ferromagnetic state including a factor of two coming from identical hopping in reverse direction is 
\begin{equation}\label{eq:FM_en}
\Delta E^{FM} = -\frac{2t^2}{U-3J_H}
\end{equation}

For an anti-ferromagnetic configuration, there are multiple exchange pathways as shown in Fig.~\ref{Fig:3}(b) and (c) of main text. The intermediate state shown in Fig.~\ref{Fig:3}(b) of main text has $S_2=1/2$ and $L_2=1$, whereas the configuration on site 1 has the following composition
\begin{equation}
\frac{1}{\sqrt{3}}\lvert S_1=3/2,L_1=0 \rangle + \sqrt{\frac{2}{3}}\lvert S_1=1/2, L_1=2 \rangle
\end{equation}
The energy gained by the anti-ferromagnetic state from this pathway is 
\begin{equation}
\Delta E^{AF}_1 = -\frac{1}{3}\frac{t^2}{U-3J_H} - \frac{2}{3} \frac{t^2}{U}
\end{equation}

Similarly, the intermediate state in the other anti-ferromagnetic exchange pathway shown in Fig.~\ref{Fig:3}(c) of main text has $S_2=1/2$, $L_2=1$ and the configuration on site 1 has the following composition
\begin{equation}
\frac{1}{\sqrt{2}}\lvert S_1=1/2,L_1=2 \rangle + \frac{1}{\sqrt{2}}\lvert S_1=1/2, L_1=1 \rangle
\end{equation}
The energy gained from this pathway is
\begin{equation}
\Delta E^{AF}_2 = -\frac{1}{2}\frac{t^2}{U} - \frac{1}{2} \frac{t^2}{U+2J_H}
\end{equation}

Therefore, the total energy gained by the anti-ferromagnetic state including a factor of two coming from identical hopping in the reverse direction is
\begin{equation}\label{eq:AF_en}
\Delta E^{AF} = -\frac{2}{3}\frac{t^2}{U-3J_H} - \frac{14}{6} \frac{t^2}{U} -  \frac{t^2}{U+2J_H}
\end{equation}

It is clear from Eq.~\ref{eq:FM_en} and Eq.~\ref{eq:AF_en} that $J_H$ favors ferromagnetic superexchange. With increasing $J_H/U$ the superexchange will change from being anti-ferromagnetic to a ferromagnetic superexchange when $\Delta E^{FM} =  \Delta E^{AF}$ or $J_H/U\approx 0.19$. For a more accurate estimate this ratio, we present in Fig.~\ref{Fig:3}(d) of main text exact numerical results from a two-site calculation which shows the sign of superexchange as a function of $U/t$ and $J_H/U$. Using properly constructed ferromagnetic $S=2$ and antiferromagnetic $S=0$ states and including all intermediate states shows that the superexchange becomes ferromagnetic for $J_H/U\approx 0.1$ for realistic estimate of $U/t\approx 20$. In 4d materials, we have $J_H/U\approx 0.2$ \cite{van_der_marel_sawatzky_1988, georges_2013} which places them firmly in the ferromagnetic regime. Ignoring Hund's coupling can erroneously lead to anti-ferromagnetic superexchange.

\section{Magnetic Hamiltonian} \label{appen:B}
In this section, we present a detailed derivation of the spin-orbital Hamiltonian in Eq.~6 of the main text.  In the atomic limit with no SOC, the $d^4$ Mott insulator has $L_i=1$ and $S_i=1$ at each site. For two sites, the ground state with $d^4-d^4$ configuration is a direct product state 
\begin{equation}
\lvert \Psi_{GS} \rangle =  \lvert S_1=1 \rangle \otimes \lvert L_1=1 \rangle \otimes \lvert S_2=1\rangle \otimes \lvert L_2=1\rangle
\end{equation}
which can give rise to total $L=2,1 \text{ or } 0$ and total $S=2,1 \text{ or } 0$, all of which are degenerate with energy $E_{0}$ (see Eq.~\ref{eq:E0}). From second order perturbation theory, the magnetic exchange term that captures the correction to the atomic ground state energy has the form  
\begin{equation}\label{eq:exchange}
\tilde{H'}= H_{hop}\left[\sum_{n\alpha}\frac{\lvert\psi_{n\alpha}\rangle\langle \psi_{n\alpha}\rvert}{E_0-E_n}\right]H_{hop}
\end{equation}

where $H_{hop}$ is the kinetic energy described in Eq.~\ref{eq:kinetic} and $\lvert \psi_{n\alpha}\rangle$ is the intermediate exited atomic state with $d^3-d^5$ configuration and energy $E_n$. $\alpha$ indicates any degeneracy of the intermediate states. 

Let us now examine the excited states. The ground state for $d^3$ has $L_i=0$ and $S_i=3/2$ while $d^5$ has $L_i=1$ and $S_i=1/2$ in its ground state. So, the lowest lying excited state with energy $E_1=13U-29J_H$, which we will call $\lvert\psi_{1\alpha} \rangle$, can have total $L=1$ and total $S=2 \text{ or } 1$. Since $\lvert\psi_{1\alpha}\rangle$ provides the dominant term in Eq.~\ref{eq:exchange}, and we will only keep $\lvert\psi_{1\alpha}\rangle$ and ignore higher energy intermediate states. 


The form of $H_{hop}$ in Eq.~\ref{eq:kinetic} is invariant under rotations in both spin and orbital space. It, therefore, commutes with total $\hat{L}^2$ and total $\hat{S}^2$ operators and only connects states with the same total $L$ and total $S$. $\lvert \psi_{1\alpha}\rangle$ has $L=1$, and consequently the energy gain from the exchange term is maximized if $\lvert \Psi_{GS}\rangle$ is in the $L=1$ state. Similarly, only the $S=2$ and $S=1$ components of $\lvert \Psi_{GS}\rangle$ gains energy via virtual excitations to $\lvert \psi_{1\alpha}\rangle$. The magnetic Hamiltonian in Eq.~\ref{eq:exchange} can be written in terms of spin and orbital projection operators as
\begin{eqnarray}\label{eq:exchange2}
\tilde{H'} &\approx& - J_2 \lvert S=2\rangle\lvert L=1\rangle\langle L=1\rvert\langle S=2\rvert \nonumber \\
   && - J_1 \lvert S=1 \rangle\lvert L=1\rangle\langle L=1\rvert\langle S=1 \rvert \\
   \text{where} && \nonumber \\
   J_i&=& \sum_{\alpha} \frac{\lvert \langle \psi_{1\alpha} \rvert H_{hop} \lvert \Psi_{GS}(L=1,S=i)\rangle\rvert^2}{\lvert E_0-E_1\rvert}
\end{eqnarray}

After a rather long but straightforward algebra, we can show that 
\begin{equation}
J_2 = \frac{4t^2}{U-3J_H} \qquad J_1 = \frac{4t^2}{3(U-3J_H)}
\end{equation}
Therefore, magnetic Hamiltonian in Eq.~\ref{eq:exchange2} can be written as
\begin{equation}\label{eq:exchange3}
\tilde{H} \approx -J_{FM}\mathbf{S_1}\cdot\mathbf{S_2}\mathcal{P}(\mathbf{L_1}+\mathbf{L_2}=1)
\end{equation}
where $J_{FM}=4t^2/(U-3J_H)$ and $\mathcal{P}$ is the same as $\lvert L=1\rangle\langle L=1\rvert$ which projects the total $L$ of the two sites to $L=1$, and it has the form
\begin{equation}
\mathcal{P}(\mathbf{L_1}+\mathbf{L_2}=1) = \frac{(1-\mathbf{L_1}.\mathbf{L_2})(2+\mathbf{L_1}.\mathbf{L_2})}{2}
\end{equation}

We can add to Eq.~\ref{eq:exchange3} the spin-orbit coupling term and generalize to a lattice in order to obtain the desired spin-orbital Hamiltonian
\begin{equation}\label{eq:spinH}
\tilde{H} =-\frac{J_{FM}}{2}\sum_{\langle ij\rangle}\mathbf{S}_i\cdot\mathbf{S}_j \mathcal{P}(\mathbf{L}_i+\mathbf{L}_j=1)
+ \frac{\lambda}{2}\sum_i\mathbf{L}_i\cdot\mathbf{S}_i
\end{equation}


\section{Resonant X-ray Scattering} \label{appen:C}

In this section, we describe the general theory of resonant x-ray scattering (RXS) that we have used to calculate the results shown in Figs.~\ref{Fig:4} of the main text. The starting point is the scattering amplitude. Within second order perturbation theory and the dipole approximation, the resonant scattering amplitude has the following form  \citep{fink_2013}
\be
\Delta f(\omega) \propto \sum_{n} \frac{ \bra{\Psi_{GS}} (\vec{e}'.\vec{D})^\dagger \ketbra{\psi_n}{\psi_n}\vec{e}.\vec{D}  \ket{\Psi_{GS}} }{E_n - E_G - \hbar\omega - i\Gamma_n}
\label{eq:amplitude}
\ee
where $\ket{\Psi_{GS}}$ is the ground state with energy $E_G$ and $\ket{\psi_n}$ is an excited state with energy $E_n$. $\Gamma_n$ corresponds to the inverse lifetime of the particular excited state $\ket{\psi_n}$ and $\vec{e} (\vec{e}')$ is the polarization of the incoming(outgoing) X-ray photon. 

It is convenient to write the dipole operator $\vec{D}$ in second quantized form to facilitate calculation of the matrix elements in the numerator of Eq.~(\ref{eq:amplitude}). For the $L_{2(3)}$ edge, absorbing a photon promotes a core $2p$ electron to the valence $d$ shell
\be
\vec{e}.\vec{D} \approx \vec{e} \cdot \hat{\vec{r}} =
\sum_{\alpha\beta\sigma} \vec{e} \cdot \bra{d_{\alpha}} \hat{\vec{r}} \ket{p_{\beta}}
d^{\dagger}_{\alpha\sigma} p^{\phd}_{\beta\sigma} + \mbox{h.c.}
\ee
where the $\bra{d_{\alpha}} \hat{\vec{r}} \ket{p_{\beta}} \equiv R_{\alpha\beta}$ are easily determined by symmetry and tabulated 
\be
R \propto
\left( \begin{array}{c|ccc}
 & \ket{p_x} & \ket{p_y} & \ket{p_z} \\
\hline
\bra{d_{yz}} & 0 & \hat{z} & \hat{y} \\
\bra{d_{zx}} & \hat{z} & 0 & \hat{x} \\
\bra{d_{xy}} & \hat{y} & \hat{x} & 0
\end{array} \right)
\qquad 
\ee
The proportionality constant depends on fine details of the atomic states, but symmetry dictates that it should be the same for all combinations of $p$ and $d$ orbitals. Hence it is an overall constant which we hereafter ignore.

\emph{Free ion approximation}: A common practice in calculating RXS matrix elements is to approximate the scattering site as a free ion \cite{kim_2009,fink_2013}. This usually gives a good description for Mott insulators where the scattering amplitude is primarily determined by local properties. The effect of the lattice comes only through the geometrical structure factor. Within the free ion approximation, the ground state in Eq.~\ref{eq:amplitude} is replaced by the atomic ground state and the excited states are replaced by atomic excited states. 

\emph{Non-local effects}: When interaction between different sites have significant effect on local properties, as in the case of the $d^4$ ferromagnetic Mott insulator, the free ion approximation breaks down. Substituting the ground state in Eq.~\ref{eq:amplitude} by the atomic ground state is no longer a good approximation. However, it turns out the excited states can still be substituted by the atomic exited states because the core hole generates an additional binding energy for the excited electron \cite{hannon_1988,fink_2013}. 

To include the non-local effects correctly, we need to write the ground state as a direct product of states defined only on the scattering site $\ket{\psi_n}$ and states defined on the rest of the lattice $\ket{\phi_n}$
\begin{equation}
\ket{\Psi_{GS}} =\sum_{p q}a_{pq}\ket{\psi_p}\ket{\phi_q}
\end{equation}
Substituting this into Eq.~\ref{eq:amplitude}, the matrix element in the numerator becomes
\begin{eqnarray}
&&\bra{\Psi_{GS}} (\vec{e}'.\vec{D})^\dagger\ketbra{\psi_n}{\psi_n}\vec{e}.\vec{D}  \ket{\Psi_{GS}} \nonumber \\ 
 && \qquad = \sum_{p qrs}a^*_{rs} a_{pq}\bra{\psi_r}(\vec{e}'.\vec{D})^\dagger \ketbra{\psi_n}{\psi_n}\vec{e}.\vec{D}\ket{\psi_p}\braket{\phi_s}{\phi_q} \nonumber \\
  &&  \qquad = \sum_{pr} \rho_{pr}\bra{\psi_r}(\vec{e}'.\vec{D})^\dagger \ketbra{\psi_n}{\psi_n}\vec{e}.\vec{D}\ket{\psi_p} \nonumber \\
  && \qquad =  \mbox{Tr}\left[\rho (\vec{e}'.\vec{D})^\dagger \ketbra{\psi_n}{\psi_n}\vec{e}.\vec{D} \right]
\end{eqnarray}
where $\rho_{pr}=\sum_{q}a^*_{rq}a_{pq}$ is the reduced density matrix at the scattering site. Finally, we get the desired expression for the resonant scattering amplitude with non-local effects included correctly
\begin{equation}
\Delta f(\omega) \propto \mbox{Tr} \left[ \rho \sum_{n}\frac{ (\mathbf{e'}.\mathbf{D})^\dagger \lvert \psi_n\rangle\langle \psi_n \rvert \mathbf{e}.\mathbf{D} }{ E_n - E_G - \hbar \omega -i\Gamma_n } \right]
\end{equation}
The influence of the lattice on the scattering site through the super-exchange interaction is crucial in understanding the ferromagnetic $d^4$ Mott insulator. 

In Fig.~\ref{Fig:4}(a) and (b) of main text, we show the magnetic ($\sigma-\pi$) scattering cross section of the ferromagnetic $d^4$ Mott insulator. Both $L_2$ and $L_3$ edges are significantly enhanced. The non-magnetic insulator for larger $\lambda$ does not exhibit any magnetic scattering and, therefore, the ($\sigma-\pi$) RXS can be used as a diagnostic for distinguishing the two magnetic phases. In our calculation, the intermediate states contributing to the resonant enhancement are the $J=1/2$ and $J=3/2$ states in the $d^5$ configuration. We include all the intermediate states assuming that the energy splitting between them is much smaller than the inverse life time $\Gamma_n$ of the excited states. If we include only the lower energy $J=1/2$ states, we find that the $L_3$ edge is completely suppressed while the $L_2$ edge is enhanced. This is the exact opposite of the case in iridates where the $L_2$ edge is completely suppressed \cite{kim_2009}.

\section{Different choices for the hopping matrix} \label{appen:D}
The Hamiltonian in Eq(1) of the main text assumes that the hopping matrix elements are diagonal and symmetric in orbital space. This is a simplification because hopping matrix elements in real materials can be strongly dependent on orbitals and bond angles. For the case of 180$^0$ bond angles relevant for our two site analysis, only two of the three $t_{2g}$ orbitals contribute to hopping \cite{harris_2004}. It raises two important questions: 1) Will the sign of super-exchange interaction change if we use a more realistic model? 2) Does our simple model capture the low energy physics properly? We address both of these issues here. 

\emph{Super-exchange:} We have repeated the exact diagonalization calculations for the two site system with realistic hopping terms - only two $t_{2g}$ orbitals contributing to hopping. In the absence of spin-orbit coupling ($\lambda$=0) the ground state has S=2 which proves beyond doubt that the super-exchange is always ferromagnetic. 

\begin{figure}[t]
\includegraphics[width=0.48\textwidth]{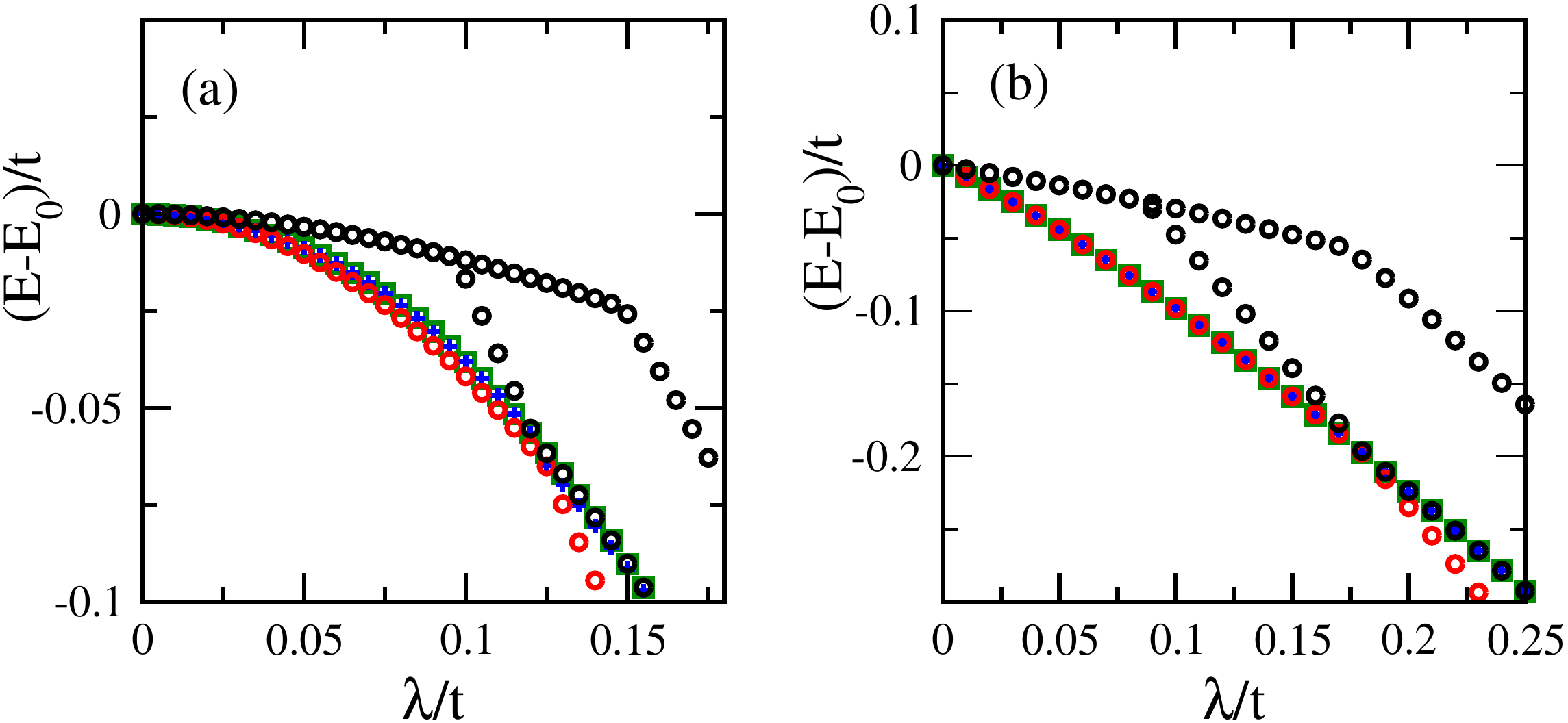}
\caption{Evolution of lowest lying energy levels of two site system as a function of $\lambda$: a) Realistic model with only two $t_{2g}$ orbitals contributing to hoping and b) Simple model with hopping diagonal and symmetric in orbital space. Energy levels are obtained by exact diagonalization. We used U=20t and JH=0.2U. E$_0$ is the energy of ground state at $\lambda$=0 in each case. Note that kinks in the evolution of energy levels are just level crossings with the higher energy state removed for clarity.}
\label{Fig:5}
\end{figure}

\emph{Low energy physics:} Fig.~\ref{Fig:5} (a) and (b) show the evolution of the lowest lying eigenvalues as a function of $\lambda$ for the two site system with realistic and simplified Hamiltonians respectively. The ground state in the realistic case is non-degenerate, however it is clear that the relevant low energy levels consists of three nearly degenerate states and a non-degenerate state which cross at around $\lambda$=0.12t. At a qualitative level, a very similar evolution of energy levels is realized in the simplified model. The only difference is that the three nearly degenerate levels become exactly degenerate for hopping which is diagonal and symmetric in orbital space. This simplification is reasonable because the energy splitting of the nearly degenerate levels is smaller than any other energy scale of the problem. Since the goal of our paper is to illustrate the existence of a new ferromagnetic state, we have chosen to work with the simpler model.

\bibliography{references}

\begin{thebibliography}{29}%
\makeatletter
\providecommand \@ifxundefined [1]{%
 \@ifx{#1\undefined}
}%
\providecommand \@ifnum [1]{%
 \ifnum #1\expandafter \@firstoftwo
 \else \expandafter \@secondoftwo
 \fi
}%
\providecommand \@ifx [1]{%
 \ifx #1\expandafter \@firstoftwo
 \else \expandafter \@secondoftwo
 \fi
}%
\providecommand \natexlab [1]{#1}%
\providecommand \enquote  [1]{``#1''}%
\providecommand \bibnamefont  [1]{#1}%
\providecommand \bibfnamefont [1]{#1}%
\providecommand \citenamefont [1]{#1}%
\providecommand \href@noop [0]{\@secondoftwo}%
\providecommand \href [0]{\begingroup \@sanitize@url \@href}%
\providecommand \@href[1]{\@@startlink{#1}\@@href}%
\providecommand \@@href[1]{\endgroup#1\@@endlink}%
\providecommand \@sanitize@url [0]{\catcode `\\12\catcode `\$12\catcode
  `\&12\catcode `\#12\catcode `\^12\catcode `\_12\catcode `\%12\relax}%
\providecommand \@@startlink[1]{}%
\providecommand \@@endlink[0]{}%
\providecommand \url  [0]{\begingroup\@sanitize@url \@url }%
\providecommand \@url [1]{\endgroup\@href {#1}{\urlprefix }}%
\providecommand \urlprefix  [0]{URL }%
\providecommand \Eprint [0]{\href }%
\providecommand \doibase [0]{http://dx.doi.org/}%
\providecommand \selectlanguage [0]{\@gobble}%
\providecommand \bibinfo  [0]{\@secondoftwo}%
\providecommand \bibfield  [0]{\@secondoftwo}%
\providecommand \translation [1]{[#1]}%
\providecommand \BibitemOpen [0]{}%
\providecommand \bibitemStop [0]{}%
\providecommand \bibitemNoStop [0]{.\EOS\space}%
\providecommand \EOS [0]{\spacefactor3000\relax}%
\providecommand \BibitemShut  [1]{\csname bibitem#1\endcsname}%
\let\auto@bib@innerbib\@empty
\bibitem [{\citenamefont {Lee}\ \emph {et~al.}(2006)\citenamefont {Lee},
  \citenamefont {Nagaosa},\ and\ \citenamefont {Wen}}]{highTc_RMP_2006}%
  \BibitemOpen
  \bibfield  {author} {\bibinfo {author} {\bibfnamefont {P.~A.}\ \bibnamefont
  {Lee}}, \bibinfo {author} {\bibfnamefont {N.}~\bibnamefont {Nagaosa}}, \ and\
  \bibinfo {author} {\bibfnamefont {X.-G.}\ \bibnamefont {Wen}},\ }\href
  {http://link.aps.org/doi/10.1103/RevModPhys.78.17} {\bibfield  {journal}
  {\bibinfo  {journal} {Rev. Mod. Phys.}\ }\textbf {\bibinfo {volume} {78}},\
  \bibinfo {pages} {17} (\bibinfo {year} {2006})}\BibitemShut {NoStop}%
\bibitem [{\citenamefont {Salamon}\ and\ \citenamefont
  {Jaime}(2001)}]{salamon_2001}%
  \BibitemOpen
  \bibfield  {author} {\bibinfo {author} {\bibfnamefont {M.~B.}\ \bibnamefont
  {Salamon}}\ and\ \bibinfo {author} {\bibfnamefont {M.}~\bibnamefont
  {Jaime}},\ }\href {http://link.aps.org/doi/10.1103/RevModPhys.73.583}
  {\bibfield  {journal} {\bibinfo  {journal} {Rev. Mod. Phys.}\ }\textbf
  {\bibinfo {volume} {73}},\ \bibinfo {pages} {583} (\bibinfo {year}
  {2001})}\BibitemShut {NoStop}%
\bibitem [{\citenamefont {Hasan}\ and\ \citenamefont
  {Kane}(2010)}]{hasan_2010}%
  \BibitemOpen
  \bibfield  {author} {\bibinfo {author} {\bibfnamefont {M.~Z.}\ \bibnamefont
  {Hasan}}\ and\ \bibinfo {author} {\bibfnamefont {C.~L.}\ \bibnamefont
  {Kane}},\ }\href {http://link.aps.org/doi/10.1103/RevModPhys.82.3045}
  {\bibfield  {journal} {\bibinfo  {journal} {Rev. Mod. Phys.}\ }\textbf
  {\bibinfo {volume} {82}},\ \bibinfo {pages} {3045} (\bibinfo {year}
  {2010})}\BibitemShut {NoStop}%
\bibitem [{\citenamefont {Pesin}\ and\ \citenamefont
  {Balents}(2010)}]{pesin_2010}%
  \BibitemOpen
  \bibfield  {author} {\bibinfo {author} {\bibfnamefont {D.}~\bibnamefont
  {Pesin}}\ and\ \bibinfo {author} {\bibfnamefont {L.}~\bibnamefont
  {Balents}},\ }\href {http://dx.doi.org/10.1038/nphys1606} {\bibfield
  {journal} {\bibinfo  {journal} {Nat Phys}\ }\textbf {\bibinfo {volume} {6}},\
  \bibinfo {pages} {376} (\bibinfo {year} {2010})}\BibitemShut {NoStop}%
\bibitem [{\citenamefont {Wan}\ \emph {et~al.}(2011)\citenamefont {Wan},
  \citenamefont {Turner}, \citenamefont {Vishwanath},\ and\ \citenamefont
  {Savrasov}}]{vishwanath_2011}%
  \BibitemOpen
  \bibfield  {author} {\bibinfo {author} {\bibfnamefont {X.}~\bibnamefont
  {Wan}}, \bibinfo {author} {\bibfnamefont {A.~M.}\ \bibnamefont {Turner}},
  \bibinfo {author} {\bibfnamefont {A.}~\bibnamefont {Vishwanath}}, \ and\
  \bibinfo {author} {\bibfnamefont {S.~Y.}\ \bibnamefont {Savrasov}},\ }\href
  {http://link.aps.org/doi/10.1103/PhysRevB.83.205101} {\bibfield  {journal}
  {\bibinfo  {journal} {Phys. Rev. B}\ }\textbf {\bibinfo {volume} {83}},\
  \bibinfo {pages} {205101} (\bibinfo {year} {2011})}\BibitemShut {NoStop}%
\bibitem [{\citenamefont {Kim}\ \emph {et~al.}(2009)\citenamefont {Kim},
  \citenamefont {Ohsumi}, \citenamefont {Komesu}, \citenamefont {Sakai},
  \citenamefont {Morita}, \citenamefont {Takagi},\ and\ \citenamefont
  {Arima}}]{kim_2009}%
  \BibitemOpen
  \bibfield  {author} {\bibinfo {author} {\bibfnamefont {B.~J.}\ \bibnamefont
  {Kim}}, \bibinfo {author} {\bibfnamefont {H.}~\bibnamefont {Ohsumi}},
  \bibinfo {author} {\bibfnamefont {T.}~\bibnamefont {Komesu}}, \bibinfo
  {author} {\bibfnamefont {S.}~\bibnamefont {Sakai}}, \bibinfo {author}
  {\bibfnamefont {T.}~\bibnamefont {Morita}}, \bibinfo {author} {\bibfnamefont
  {H.}~\bibnamefont {Takagi}}, \ and\ \bibinfo {author} {\bibfnamefont
  {T.}~\bibnamefont {Arima}},\ }\href
  {http://www.sciencemag.org/content/323/5919/1329} {\bibfield  {journal}
  {\bibinfo  {journal} {Science}\ }\textbf {\bibinfo {volume} {323}},\ \bibinfo
  {pages} {1329} (\bibinfo {year} {2009})}\BibitemShut {NoStop}%
\bibitem [{\citenamefont {Ueda}\ \emph {et~al.}(2012)\citenamefont {Ueda},
  \citenamefont {Fujioka}, \citenamefont {Takahashi}, \citenamefont {Suzuki},
  \citenamefont {Ishiwata}, \citenamefont {Taguchi},\ and\ \citenamefont
  {Tokura}}]{ueda_2012}%
  \BibitemOpen
  \bibfield  {author} {\bibinfo {author} {\bibfnamefont {K.}~\bibnamefont
  {Ueda}}, \bibinfo {author} {\bibfnamefont {J.}~\bibnamefont {Fujioka}},
  \bibinfo {author} {\bibfnamefont {Y.}~\bibnamefont {Takahashi}}, \bibinfo
  {author} {\bibfnamefont {T.}~\bibnamefont {Suzuki}}, \bibinfo {author}
  {\bibfnamefont {S.}~\bibnamefont {Ishiwata}}, \bibinfo {author}
  {\bibfnamefont {Y.}~\bibnamefont {Taguchi}}, \ and\ \bibinfo {author}
  {\bibfnamefont {Y.}~\bibnamefont {Tokura}},\ }\href
  {http://link.aps.org/doi/10.1103/PhysRevLett.109.136402} {\bibfield
  {journal} {\bibinfo  {journal} {Phys. Rev. Lett.}\ }\textbf {\bibinfo
  {volume} {109}},\ \bibinfo {pages} {136402} (\bibinfo {year}
  {2012})}\BibitemShut {NoStop}%
\bibitem [{\citenamefont {Chen}\ \emph {et~al.}(2010)\citenamefont {Chen},
  \citenamefont {Pereira},\ and\ \citenamefont {Balents}}]{chen_2010}%
  \BibitemOpen
  \bibfield  {author} {\bibinfo {author} {\bibfnamefont {G.}~\bibnamefont
  {Chen}}, \bibinfo {author} {\bibfnamefont {R.}~\bibnamefont {Pereira}}, \
  and\ \bibinfo {author} {\bibfnamefont {L.}~\bibnamefont {Balents}},\ }\href
  {http://link.aps.org/doi/10.1103/PhysRevB.82.174440} {\bibfield  {journal}
  {\bibinfo  {journal} {Phys. Rev. B}\ }\textbf {\bibinfo {volume} {82}},\
  \bibinfo {pages} {174440} (\bibinfo {year} {2010})}\BibitemShut {NoStop}%
\bibitem [{\citenamefont {Chen}\ and\ \citenamefont
  {Balents}(2011)}]{chen_2011}%
  \BibitemOpen
  \bibfield  {author} {\bibinfo {author} {\bibfnamefont {G.}~\bibnamefont
  {Chen}}\ and\ \bibinfo {author} {\bibfnamefont {L.}~\bibnamefont {Balents}},\
  }\href {http://link.aps.org/doi/10.1103/PhysRevB.84.094420} {\bibfield
  {journal} {\bibinfo  {journal} {Phys. Rev. B}\ }\textbf {\bibinfo {volume}
  {84}},\ \bibinfo {pages} {094420} (\bibinfo {year} {2011})}\BibitemShut
  {NoStop}%
\bibitem [{\citenamefont {Meetei}\ \emph {et~al.}(2013)\citenamefont {Meetei},
  \citenamefont {Erten}, \citenamefont {Randeria}, \citenamefont {Trivedi},\
  and\ \citenamefont {Woodward}}]{meetei_2013}%
  \BibitemOpen
  \bibfield  {author} {\bibinfo {author} {\bibfnamefont {O.~N.}\ \bibnamefont
  {Meetei}}, \bibinfo {author} {\bibfnamefont {O.}~\bibnamefont {Erten}},
  \bibinfo {author} {\bibfnamefont {M.}~\bibnamefont {Randeria}}, \bibinfo
  {author} {\bibfnamefont {N.}~\bibnamefont {Trivedi}}, \ and\ \bibinfo
  {author} {\bibfnamefont {P.}~\bibnamefont {Woodward}},\ }\href
  {http://link.aps.org/doi/10.1103/PhysRevLett.110.087203} {\bibfield
  {journal} {\bibinfo  {journal} {Phys. Rev. Lett.}\ }\textbf {\bibinfo
  {volume} {110}},\ \bibinfo {pages} {087203} (\bibinfo {year}
  {2013})}\BibitemShut {NoStop}%
\bibitem [{\citenamefont {Khaliullin}(2013)}]{khaliullin_2013}%
  \BibitemOpen
  \bibfield  {author} {\bibinfo {author} {\bibfnamefont {G.}~\bibnamefont
  {Khaliullin}},\ }\href
  {http://link.aps.org/doi/10.1103/PhysRevLett.111.197201} {\bibfield
  {journal} {\bibinfo  {journal} {Phys. Rev. Lett.}\ }\textbf {\bibinfo
  {volume} {111}},\ \bibinfo {pages} {197201} (\bibinfo {year}
  {2013})}\BibitemShut {NoStop}%
\bibitem [{foo()}]{footnote}%
  \BibitemOpen
  \href@noop {} {}\bibinfo {note} {Constrained random phase approximation
  calculations show that Ru in $d^4$ configuration has $U\approx 2.6 \, eV$
  \cite{biermann_2012}. Density functional theory (DFT) calculations show that
  Sr$_2$YRuO$_6$ has a band width $W = 8t \approx 1.1 \, eV$ \cite{mazin_1997}.
  SrYRuO$_6$ is very similar to our candidate material La$_2$ZnRuO$_6$ in
  structure. Combining the estimates of $U$ and $t$, we get $U/t\approx 20$ for
  La$_2$ZnRuO$_6$.}\BibitemShut {Stop}%
\bibitem [{\citenamefont {Lee}\ \emph {et~al.}(2003)\citenamefont {Lee},
  \citenamefont {Lee}, \citenamefont {Noh}, \citenamefont {Byun}, \citenamefont
  {Yoo}, \citenamefont {Yamaura},\ and\ \citenamefont
  {Takayama-Muromachi}}]{lee_2003}%
  \BibitemOpen
  \bibfield  {author} {\bibinfo {author} {\bibfnamefont {Y.~S.}\ \bibnamefont
  {Lee}}, \bibinfo {author} {\bibfnamefont {J.~S.}\ \bibnamefont {Lee}},
  \bibinfo {author} {\bibfnamefont {T.~W.}\ \bibnamefont {Noh}}, \bibinfo
  {author} {\bibfnamefont {D.~Y.}\ \bibnamefont {Byun}}, \bibinfo {author}
  {\bibfnamefont {K.~S.}\ \bibnamefont {Yoo}}, \bibinfo {author} {\bibfnamefont
  {K.}~\bibnamefont {Yamaura}}, \ and\ \bibinfo {author} {\bibfnamefont
  {E.}~\bibnamefont {Takayama-Muromachi}},\ }\href {\doibase
  10.1103/PhysRevB.67.113101} {\bibfield  {journal} {\bibinfo  {journal} {Phys.
  Rev. B}\ }\textbf {\bibinfo {volume} {67}},\ \bibinfo {pages} {113101}
  (\bibinfo {year} {2003})}\BibitemShut {NoStop}%
\bibitem [{\citenamefont {Vaugier}\ \emph {et~al.}(2012)\citenamefont
  {Vaugier}, \citenamefont {Jiang},\ and\ \citenamefont
  {Biermann}}]{biermann_2012}%
  \BibitemOpen
  \bibfield  {author} {\bibinfo {author} {\bibfnamefont {L.}~\bibnamefont
  {Vaugier}}, \bibinfo {author} {\bibfnamefont {H.}~\bibnamefont {Jiang}}, \
  and\ \bibinfo {author} {\bibfnamefont {S.}~\bibnamefont {Biermann}},\ }\href
  {\doibase 10.1103/PhysRevB.86.165105} {\bibfield  {journal} {\bibinfo
  {journal} {Phys. Rev. B}\ }\textbf {\bibinfo {volume} {86}},\ \bibinfo
  {pages} {165105} (\bibinfo {year} {2012})}\BibitemShut {NoStop}%
\bibitem [{\citenamefont {Imada}\ \emph {et~al.}(1998)\citenamefont {Imada},
  \citenamefont {Fujimori},\ and\ \citenamefont {Tokura}}]{imada_1998}%
  \BibitemOpen
  \bibfield  {author} {\bibinfo {author} {\bibfnamefont {M.}~\bibnamefont
  {Imada}}, \bibinfo {author} {\bibfnamefont {A.}~\bibnamefont {Fujimori}}, \
  and\ \bibinfo {author} {\bibfnamefont {Y.}~\bibnamefont {Tokura}},\ }\href
  {\doibase 10.1103/RevModPhys.70.1039} {\bibfield  {journal} {\bibinfo
  {journal} {Rev. Mod. Phys.}\ }\textbf {\bibinfo {volume} {70}},\ \bibinfo
  {pages} {1039} (\bibinfo {year} {1998})}\BibitemShut {NoStop}%
\bibitem [{\citenamefont {van~der Marel}\ and\ \citenamefont
  {Sawatzky}(1988)}]{van_der_marel_sawatzky_1988}%
  \BibitemOpen
  \bibfield  {author} {\bibinfo {author} {\bibfnamefont {D.}~\bibnamefont
  {van~der Marel}}\ and\ \bibinfo {author} {\bibfnamefont {G.~A.}\ \bibnamefont
  {Sawatzky}},\ }\href {\doibase 10.1103/PhysRevB.37.10674} {\bibfield
  {journal} {\bibinfo  {journal} {Phys. Rev. B}\ }\textbf {\bibinfo {volume}
  {37}},\ \bibinfo {pages} {10674} (\bibinfo {year} {1988})}\BibitemShut
  {NoStop}%
\bibitem [{\citenamefont {Georges}\ \emph {et~al.}(2013)\citenamefont
  {Georges}, \citenamefont {Medici},\ and\ \citenamefont
  {Mravlje}}]{georges_2013}%
  \BibitemOpen
  \bibfield  {author} {\bibinfo {author} {\bibfnamefont {A.}~\bibnamefont
  {Georges}}, \bibinfo {author} {\bibfnamefont {L.~d.}\ \bibnamefont {Medici}},
  \ and\ \bibinfo {author} {\bibfnamefont {J.}~\bibnamefont {Mravlje}},\ }\href
  {\doibase 10.1146/annurev-conmatphys-020911-125045} {\bibfield  {journal}
  {\bibinfo  {journal} {Annu. Rev. Condens. Matter Phys.}\ }\textbf {\bibinfo
  {volume} {4}},\ \bibinfo {pages} {137} (\bibinfo {year} {2013})}\BibitemShut
  {NoStop}%
\bibitem [{\citenamefont {Du}\ \emph {et~al.}(2013{\natexlab{a}})\citenamefont
  {Du}, \citenamefont {Huang},\ and\ \citenamefont {Dai}}]{du2_2013}%
  \BibitemOpen
  \bibfield  {author} {\bibinfo {author} {\bibfnamefont {L.}~\bibnamefont
  {Du}}, \bibinfo {author} {\bibfnamefont {L.}~\bibnamefont {Huang}}, \ and\
  \bibinfo {author} {\bibfnamefont {X.}~\bibnamefont {Dai}},\ }\href
  {http://link.springer.com/article/10.1140/epjb/e2013-31024-6} {\bibfield
  {journal} {\bibinfo  {journal} {Eur. Phys. J. B}\ }\textbf {\bibinfo {volume}
  {86}},\ \bibinfo {pages} {1} (\bibinfo {year}
  {2013}{\natexlab{a}})}\BibitemShut {NoStop}%
\bibitem [{\citenamefont {Du}\ \emph {et~al.}(2013{\natexlab{b}})\citenamefont
  {Du}, \citenamefont {Sheng}, \citenamefont {Weng},\ and\ \citenamefont
  {Dai}}]{du1_2013}%
  \BibitemOpen
  \bibfield  {author} {\bibinfo {author} {\bibfnamefont {L.}~\bibnamefont
  {Du}}, \bibinfo {author} {\bibfnamefont {X.}~\bibnamefont {Sheng}}, \bibinfo
  {author} {\bibfnamefont {H.}~\bibnamefont {Weng}}, \ and\ \bibinfo {author}
  {\bibfnamefont {X.}~\bibnamefont {Dai}},\ }\href
  {http://iopscience.iop.org/0295-5075/101/27003} {\bibfield  {journal}
  {\bibinfo  {journal} {Europhys. Lett.}\ }\textbf {\bibinfo {volume} {101}},\
  \bibinfo {pages} {27003} (\bibinfo {year} {2013}{\natexlab{b}})}\BibitemShut
  {NoStop}%
\bibitem [{\citenamefont {Mazin}\ and\ \citenamefont
  {Singh}(1997)}]{mazin_1997}%
  \BibitemOpen
  \bibfield  {author} {\bibinfo {author} {\bibfnamefont {I.~I.}\ \bibnamefont
  {Mazin}}\ and\ \bibinfo {author} {\bibfnamefont {D.~J.}\ \bibnamefont
  {Singh}},\ }\href {http://link.aps.org/doi/10.1103/PhysRevB.56.2556}
  {\bibfield  {journal} {\bibinfo  {journal} {Phys. Rev. B}\ }\textbf {\bibinfo
  {volume} {56}},\ \bibinfo {pages} {2556} (\bibinfo {year}
  {1997})}\BibitemShut {NoStop}%
\bibitem [{\citenamefont {Fink}\ \emph {et~al.}(2013)\citenamefont {Fink},
  \citenamefont {Schierle}, \citenamefont {Weschke},\ and\ \citenamefont
  {Geck}}]{fink_2013}%
  \BibitemOpen
  \bibfield  {author} {\bibinfo {author} {\bibfnamefont {J.}~\bibnamefont
  {Fink}}, \bibinfo {author} {\bibfnamefont {E.}~\bibnamefont {Schierle}},
  \bibinfo {author} {\bibfnamefont {E.}~\bibnamefont {Weschke}}, \ and\
  \bibinfo {author} {\bibfnamefont {J.}~\bibnamefont {Geck}},\ }\href
  {http://iopscience.iop.org/0034-4885/76/5/056502} {\bibfield  {journal}
  {\bibinfo  {journal} {Rep. Prog. Phys.}\ }\textbf {\bibinfo {volume} {76}},\
  \bibinfo {pages} {056502} (\bibinfo {year} {2013})}\BibitemShut {NoStop}%
\bibitem [{\citenamefont {Sachdev}\ and\ \citenamefont
  {Bhatt}(1990)}]{sachdev_1990}%
  \BibitemOpen
  \bibfield  {author} {\bibinfo {author} {\bibfnamefont {S.}~\bibnamefont
  {Sachdev}}\ and\ \bibinfo {author} {\bibfnamefont {R.~N.}\ \bibnamefont
  {Bhatt}},\ }\href {http://link.aps.org/doi/10.1103/PhysRevB.41.9323}
  {\bibfield  {journal} {\bibinfo  {journal} {Phys. Rev. B}\ }\textbf {\bibinfo
  {volume} {41}},\ \bibinfo {pages} {9323} (\bibinfo {year}
  {1990})}\BibitemShut {NoStop}%
\bibitem [{\citenamefont {Giamarchi}\ \emph {et~al.}(2008)\citenamefont
  {Giamarchi}, \citenamefont {R\~{u}egg},\ and\ \citenamefont
  {Tchernyshyov}}]{giamarchi_2008}%
  \BibitemOpen
  \bibfield  {author} {\bibinfo {author} {\bibfnamefont {T.}~\bibnamefont
  {Giamarchi}}, \bibinfo {author} {\bibfnamefont {C.}~\bibnamefont
  {R\~{u}egg}}, \ and\ \bibinfo {author} {\bibfnamefont {O.}~\bibnamefont
  {Tchernyshyov}},\ }\href
  {http://www.nature.com/nphys/journal/v4/n3/abs/nphys893.html} {\bibfield
  {journal} {\bibinfo  {journal} {Nature Phys.}\ }\textbf {\bibinfo {volume}
  {4}},\ \bibinfo {pages} {198} (\bibinfo {year} {2008})}\BibitemShut {NoStop}%
\bibitem [{\citenamefont {Dass}\ \emph {et~al.}(2004)\citenamefont {Dass},
  \citenamefont {Yan},\ and\ \citenamefont {Goodenough}}]{dass_2004}%
  \BibitemOpen
  \bibfield  {author} {\bibinfo {author} {\bibfnamefont {R.~I.}\ \bibnamefont
  {Dass}}, \bibinfo {author} {\bibfnamefont {J.-Q.}\ \bibnamefont {Yan}}, \
  and\ \bibinfo {author} {\bibfnamefont {J.~B.}\ \bibnamefont {Goodenough}},\
  }\href {http://link.aps.org/doi/10.1103/PhysRevB.69.094416} {\bibfield
  {journal} {\bibinfo  {journal} {Phys. Rev. B}\ }\textbf {\bibinfo {volume}
  {69}},\ \bibinfo {pages} {094416} (\bibinfo {year} {2004})}\BibitemShut
  {NoStop}%
\bibitem [{\citenamefont {Yoshii}\ \emph {et~al.}(2006)\citenamefont {Yoshii},
  \citenamefont {Ikeda},\ and\ \citenamefont {Mizumaki}}]{yoshii_2006}%
  \BibitemOpen
  \bibfield  {author} {\bibinfo {author} {\bibfnamefont {K.}~\bibnamefont
  {Yoshii}}, \bibinfo {author} {\bibfnamefont {N.}~\bibnamefont {Ikeda}}, \
  and\ \bibinfo {author} {\bibfnamefont {M.}~\bibnamefont {Mizumaki}},\ }\href
  {\doibase 10.1002/pssa.200669549} {\bibfield  {journal} {\bibinfo  {journal}
  {physica status solidi (a)}\ }\textbf {\bibinfo {volume} {203}},\ \bibinfo
  {pages} {2812–2817} (\bibinfo {year} {2006})}\BibitemShut {NoStop}%
\bibitem [{\citenamefont {de’ Medici}\ \emph {et~al.}(2011)\citenamefont
  {de’ Medici}, \citenamefont {Mravlje},\ and\ \citenamefont
  {Georges}}]{antoine_2011}%
  \BibitemOpen
  \bibfield  {author} {\bibinfo {author} {\bibfnamefont {L.}~\bibnamefont
  {de’ Medici}}, \bibinfo {author} {\bibfnamefont {J.}~\bibnamefont
  {Mravlje}}, \ and\ \bibinfo {author} {\bibfnamefont {A.}~\bibnamefont
  {Georges}},\ }\href {\doibase 10.1103/PhysRevLett.107.256401} {\bibfield
  {journal} {\bibinfo  {journal} {Phys. Rev. Lett.}\ }\textbf {\bibinfo
  {volume} {107}},\ \bibinfo {pages} {256401} (\bibinfo {year}
  {2011})}\BibitemShut {NoStop}%
\bibitem [{\citenamefont {Li}\ \emph {et~al.}(2014)\citenamefont {Li},
  \citenamefont {Lieb},\ and\ \citenamefont {Wu}}]{li_2014}%
  \BibitemOpen
  \bibfield  {author} {\bibinfo {author} {\bibfnamefont {Y.}~\bibnamefont
  {Li}}, \bibinfo {author} {\bibfnamefont {E.~H.}\ \bibnamefont {Lieb}}, \ and\
  \bibinfo {author} {\bibfnamefont {C.}~\bibnamefont {Wu}},\ }\href {\doibase
  10.1103/PhysRevLett.112.217201} {\bibfield  {journal} {\bibinfo  {journal}
  {Phys. Rev. Lett.}\ }\textbf {\bibinfo {volume} {112}},\ \bibinfo {pages}
  {217201} (\bibinfo {year} {2014})}\BibitemShut {NoStop}%
\bibitem [{\citenamefont {Hannon}\ \emph {et~al.}(1988)\citenamefont {Hannon},
  \citenamefont {Trammell}, \citenamefont {Blume},\ and\ \citenamefont
  {Gibbs}}]{hannon_1988}%
  \BibitemOpen
  \bibfield  {author} {\bibinfo {author} {\bibfnamefont {J.~P.}\ \bibnamefont
  {Hannon}}, \bibinfo {author} {\bibfnamefont {G.~T.}\ \bibnamefont
  {Trammell}}, \bibinfo {author} {\bibfnamefont {M.}~\bibnamefont {Blume}}, \
  and\ \bibinfo {author} {\bibfnamefont {D.}~\bibnamefont {Gibbs}},\ }\href
  {http://link.aps.org/doi/10.1103/PhysRevLett.61.1245} {\bibfield  {journal}
  {\bibinfo  {journal} {Phys. Rev. Lett.}\ }\textbf {\bibinfo {volume} {61}},\
  \bibinfo {pages} {1245} (\bibinfo {year} {1988})}\BibitemShut {NoStop}%
\bibitem [{\citenamefont {Harris}\ \emph {et~al.}(2004)\citenamefont {Harris},
  \citenamefont {Yildirim}, \citenamefont {Aharony}, \citenamefont
  {Entin-Wohlman},\ and\ \citenamefont {Korenblit}}]{harris_2004}%
  \BibitemOpen
  \bibfield  {author} {\bibinfo {author} {\bibfnamefont {A.~B.}\ \bibnamefont
  {Harris}}, \bibinfo {author} {\bibfnamefont {T.}~\bibnamefont {Yildirim}},
  \bibinfo {author} {\bibfnamefont {A.}~\bibnamefont {Aharony}}, \bibinfo
  {author} {\bibfnamefont {O.}~\bibnamefont {Entin-Wohlman}}, \ and\ \bibinfo
  {author} {\bibfnamefont {I.~Y.}\ \bibnamefont {Korenblit}},\ }\href
  {http://link.aps.org/doi/10.1103/PhysRevB.69.035107} {\bibfield  {journal}
  {\bibinfo  {journal} {Phys. Rev. B}\ }\textbf {\bibinfo {volume} {69}},\
  \bibinfo {pages} {035107} (\bibinfo {year} {2004})}\BibitemShut {NoStop}%
\end{thebibliography}%
\bibliographystyle{apsrev4-1}
 
\end{document}